\documentclass[a4paper,aps,pra,twocolumn,preprintnumbers,amsmath,amssymb]{revtex4}
\usepackage{bbm}
\usepackage{amsmath}
\usepackage{verbatim}
\usepackage{graphicx}
\bibliography{References}
\begin{document}

\title{Quantum Information at the Interface of Light with Atomic Ensembles and Micromechanical Oscillators}

\author{Christine A. Muschik$^1$, Hanna Krauter$^2$, Klemens Hammerer$^3$, and Eugene S. Polzik$^{2}$}

\affiliation{ $^1$Max-Planck--Institut f\"ur Quantenoptik,
Hans-Kopfermann-Strasse, D-85748 Garching, Germany \\
$^2$ Niels Bohr Institute, Danish Quantum Optics Center –
QUANTOP, Copenhagen University, Blegdamsvej 17, 2100 Copenhagen
Denmark.\\
$^3$Institute for Theoretical Physics, Institute for Gravitational
Physics, Leibniz University Hanover, Callinstrasse 38, D-30167
Hanover, Germany}

\begin{abstract}
This article reviews recent research towards a universal
light-matter interface. Such an interface is an important
prerequisite for long distance quantum communication, entanglement
assisted sensing and measurement, as well as for scalable photonic
quantum computation. We review the developments in light-matter
interfaces based on room temperature atomic vapors interacting
with propagating pulses via the Faraday effect. This interaction
has long been used as a tool for quantum nondemolition detections
of atomic spins via light. It was discovered recently that this
type of light-matter interaction can actually be tuned to realize
more general dynamics, enabling better performance of the
light-matter interface as well as rendering tasks possible, which
were before thought to be impractical. This includes the
realization of improved entanglement assisted and backaction
evading magnetometry approaching the Quantum Cramer-Rao limit,
quantum memory for squeezed states of light and the dissipative
generation of entanglement. A separate, but related, experiment on
entanglement assisted cold atom clock showing the Heisenberg
scaling of precision is described. We also review a possible
interface between collective atomic spins with nano- or
micromechanical oscillators, providing a link between atomic and
solid state physics approaches towards quantum information
processing.
\end{abstract}


\maketitle 

\section{Introduction}\label{Sec:Introduction}

One of the long term goals in quantum information processing is to
distribute entanglement over long distances and at large rates, in
order to serve as a resource for quantum communication protocols.
Such a quantum communication network will have to make use of a
quantum interface which allows to efficiently perform certain
primitives, such as converting light -- the natural long distance
carrier of quantum information -- to stationary quantum memories,
or creating entanglement between light and the quantum memory.
Such a device will also be necessary in order to render
architectures for photonic quantum computation scalable. For
realizing an efficient quantum interface a strong light-matter
interaction is required. Various possible approaches towards this
end are being intensely explored, ranging from single atoms inside
high finesse cavities \cite{specht_single-atom_2011}, via solid
state devices
\cite{de_riedmatten_solid-state_2008,hedges_efficient_2010,usmani_mapping_2010,saglamyurek_broadband_2011},
to ensembles of atoms interacting with light in free space
\cite{cviklinski_reversible_2008,yuan_experimental_2008,zhao_long-lived_2009,zhang_creation_2009,zhao_millisecond_2009,schnorrberger_electromagnetically_2009,reim_towards_2010,radnaev_quantum_2010,choi_entanglement_2010,hosseini_high_2011,WFJM09,JWK10,WJKRBP10}.
For comprehensive recent reviews on the various approaches based
on atomic ensembles we refer to
\cite{Hammerer2010,Sangouard2011,lvovsky_optical_2009}, see also
\cite{miller_quantum_2010}.

Here, we focus on the most recent developments in the light-matter
interface based on room temperature vapor of Cesium in glass cells
\cite{WFJM09,JWK10,WJKRBP10}. The interaction between light and
macroscopic Cesium ensembles at room temperature has been used
extensively in many different experiments over the last decade and
enabled the realization of several important quantum information
processing tasks in this system including the demonstration of a
quantum memory for light \cite{JSC04}, quantum teleportation
between light and matter \cite{SKO06} and the generation of
entanglement between two distant atomic samples using measurements
and feedback operations \cite{JKP01}. The description of these and
other experiments has been based on a quantum-nondemolition (QND)
interaction between matter and light.\\
While QND interactions combined with measurements have been proven
to be very successful, a new generation of developments and
experiments has become possible using a more general description
of the light-matter interaction. This allows not only for the
accurate description of effects which have not been taken into
account before and have therefore been treated as noise, but
allows also for the design and implementation of protocols which
go beyond the possibilities that can be realized using QND
interactions, for example the purely dissipative generation of
entanglement as described in
Sec.~\ref{Sec:EntanglementByDissipation}.\\

\section{Interaction between atomic ensembles and light}
In this section, the interaction of light with an atomic ensemble
is introduced and discussed for two-level systems
(Sec.~\ref{Sec:TwoLevel}). It is explained how it can be realized
in Cesium ensembles and how this interaction can be tuned by
varying externally controllable parameters
(Sec.~\ref{Sec:MultiLevel}). Moreover, characteristic features of
the QND-Hamiltonian and a general quadratic interaction are
highlighted and the respective input-output relations are
discussed (Sec.~\ref{Sec:IORs}).

\subsection{Light-matter interaction in a two-level model}\label{Sec:TwoLevel}
\begin{figure}
\includegraphics[width=\columnwidth]{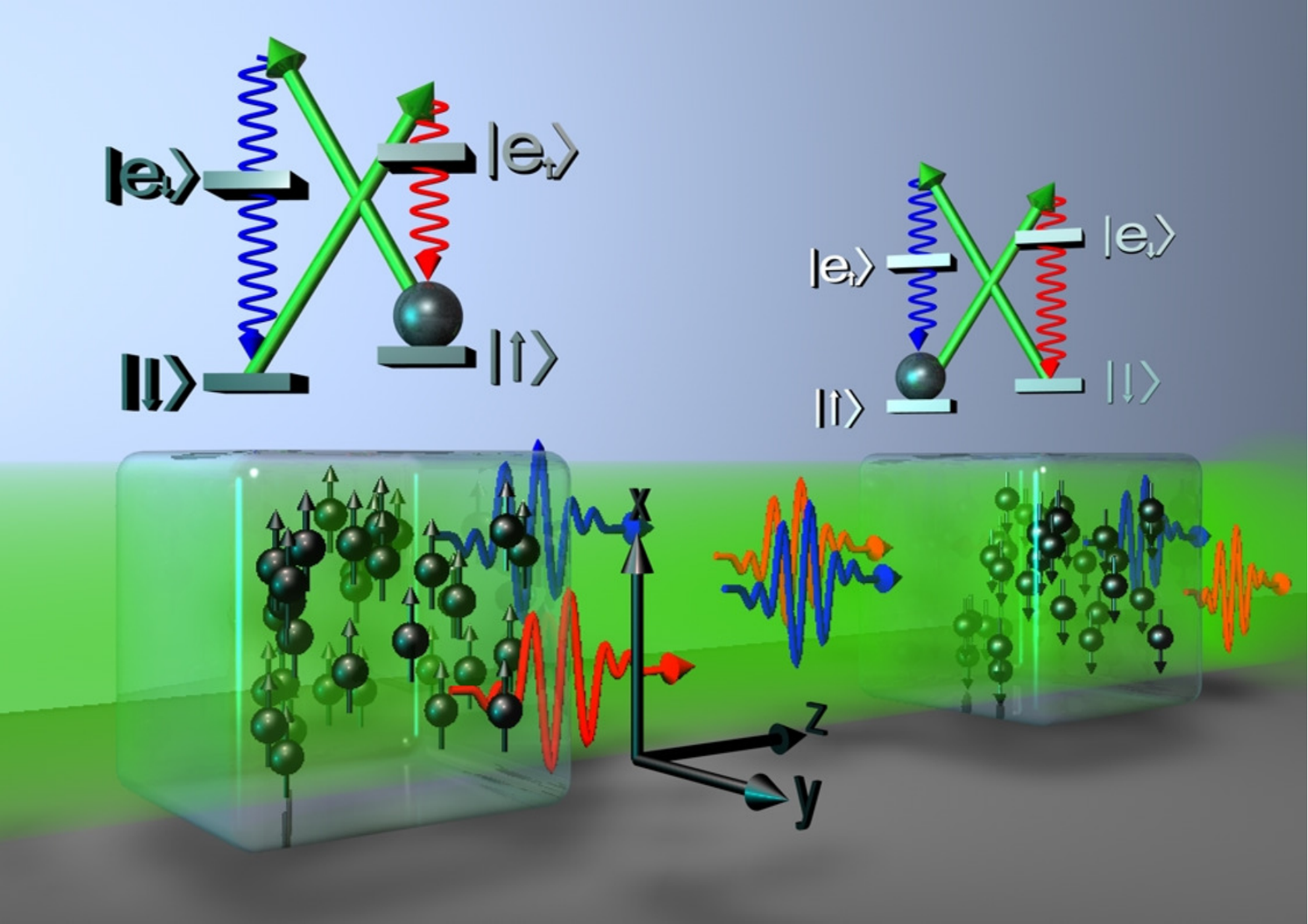}
\caption{Light-matter interaction involving a strong
$\hat{\mathbf{y}}$ polarized laser field (depicted in green) and
two atomic ensembles, which are spin polarized parallel and
antiparallel with respect to a homogeneous magnetic background
field which is oriented along $\hat{\mathbf{x}}$ and defines the
quantization axis. Atoms are assumed to posses two ground and two
excited states $|\!\!\uparrow\rangle$, $|\!\!\downarrow\rangle$
and $|e_{\uparrow}\rangle$, $|e_{\downarrow}\rangle$. The strong
off-resonant driving field induces diagonal transitions
$|\!\!\uparrow\rangle \rightarrow|e_{\downarrow}\rangle $,
$|\!\!\downarrow\rangle \rightarrow |e_{\uparrow}\rangle$ which
lead to the emission of photons in $\hat{\mathbf{x}}$ polarization
(corresponding to the transitions $|e_{\uparrow}\rangle\rightarrow
|\!\!\uparrow\rangle $, $|e_{\downarrow}\rangle\rightarrow
|\!\!\downarrow\rangle$). Due to the Zeeman splitting $\Omega$ of
the atomic ground states, photons are scattered into the upper and
lower sideband (shown in blue and red respectively) which are
centered at $\omega_L\pm \Omega$, where $\omega_L$ is the
frequency of the incident classical field.}\label{Fig:Setup}
\end{figure}
%
To start with, a simple one-dimensional two-level model involving
the ground states $|\!\!\uparrow\rangle$ and
$|\!\!\downarrow\rangle$ is considered, as illustrated in
Fig.~\ref{Fig:Setup}, which shows a $\hat{\mathbf{y}}$-polarized
laser beam propagating along $\hat{\mathbf{z}}$, and two ensembles
which are strongly spin polarized and placed in a homogeneous
magnetic field oriented along $\hat{\mathbf{x}}$. In this
subsection, we focus on the interaction of the light field with
the first ensemble which is polarized in the same direction. The
light-matter interaction is assumed to be far off-resonant and
therefore well within the dispersive regime. The excited levels
are adiabatically eliminated under the condition $\Delta\gg
\Gamma_{\mathrm{atomic}},\delta$, where $\Delta$ is the detuning,
$\Gamma_{\mathrm{atomic}}$ is the largest rate at which
transitions
$|\!\!\uparrow\rangle\leftrightarrow|\!\!\downarrow\rangle$ occur,
and $\delta$ is the Doppler width. This way, an effective
interaction involving atomic ground states only is obtained. Atoms
and light are described by means of the operators $a_A$,
$a_A^{\dag}$ and spatially localized modes $a_L(z)$,
$a_L^{\dag}(z)$ respectively. The creation operator
$a_L^{\dag}(z)$ \cite{FootnoteSpatiallyLocalizedModes} refers to a
$\hat{\mathbf{x}}$-polarized photon (the strong
$\hat{\mathbf{y}}$-polarized coherent beam is treated as classical
field) which is emitted in forward direction (emission of photons
in other directions can be included in the form of noise terms).
The operator $a_A^{\dag}=\frac{1}{\sqrt{N_A}}\sum_{i=1}^{N_A}
|\!\!\downarrow\rangle_i\langle\uparrow\!\!|$, where $N_A$ is the
number of particles in the ensemble, refers to a collective atomic
excitation. For strongly polarized samples, these operators can be
assumed to obey bosonic commutation relations
$[a_A,a_A^{\dag}]=1$, within the Holstein-Primakoff approximation
\cite{HoP40}. The action of this collective operator on the
product state
$|0\rangle_A\equiv|\!\!\uparrow_1,\uparrow_2,\dots,\uparrow_{N_A}\rangle$,
where all atoms have been initialized in $|\!\!\uparrow\rangle$,
results in the symmetric coherent superposition of all $N_A$
possible terms representing the state where one spin in the
ensemble has been flipped
$a_A^{\dag}|0\rangle_A=\frac{1}{\sqrt{N_A}}\sum_{i=1}^{N_A}|\!\uparrow_1,\dots,\downarrow_i,\dots,\uparrow_{N_A}\rangle$.\\
\\Atoms and light are assumed to interact according to a
Hamiltonian which is quadratic in the operators describing the
atomic ensemble and the light field \cite{FootnoteGaussian}. The
realization of strong nonlinearities in atomic systems would be
desirable but represents still a formidable challenge (the
realization of a cubic term would allow for universal quantum
computation if arbitrary quadratic interactions are available
\cite{LlB99}).

By means of suitable local operations, any quadratic Hamiltonian
describing the interaction of two one-mode continuous variable
systems can be parametrized by two parameters $\gamma_{s}$ and $Z$
and expressed as a sum of a passive and an active contribution
\cite{KrHGC03},
\begin{eqnarray}\label{Eq:Hamiltonian}
H_{\mathrm{int}}= \sqrt{2\gamma_{s}} \left(\mu H_P- \nu
H_A\right),
\end{eqnarray}
where $\mu=(Z+\frac{1}{Z})/2$ and $\nu=(Z-\frac{1}{Z})/2$ (compare
\cite{WFJM09}). The passive contribution $H_{P}=a_L(0)
a_A^{\dag}+H.C.$ \cite{FootnotePointlike} is energy conserving. If
a collective atomic excitation is created, a photon is
annihilated. In contrast, the active interaction $H_{A}=a_L(0)
a_A+H.C.$ corresponds to the creation (or annihilation) of atomic
and photonic excitations in pairs. The former interaction can be
understood as the interspecies analog of a beamsplitter
interaction while the latter creates entanglement and is referred
to as "squeezing interaction". The light matter interaction
considered here involves both types. The QND Hamiltonian
corresponds to the special case $|\mu|=|\nu|$, where $H_A$ and
$H_P$ contribute exactly with equal strength. In the simple
two-level model, an imbalance $|\mu| \neq|\nu|$ can arise due to
the Larmor splitting of the ground state, which leads to different
detunings $\Delta +\Omega$ and $\Delta-\Omega$ for the two photon
transitions associated with the active and the passive part of the
interaction respectively and therefore to different effective
coupling strengths and accordingly to a deviation from the QND
Hamiltonian. However, in the experimental situation considered
here, the detuning $\Delta$ is much larger than the Larmor
splitting such that this effect is negligible (in a magnetic field
of $1$ Gauss, the Zeeman shift of magnetic sublevels is about
$10^5$Hz while the detuning is on the order of $10^8$ Hz). The
non-QND character of the light-matter interaction in $^{133}$Cs
atoms is due to the fact that the levels $|\!\!\uparrow\rangle$
and $|\!\!\downarrow\rangle$ couple to several excited levels
\cite{KuMSJP05,MKP06,MKMP07,WFJM09} as described below.

\subsection{Light-matter interaction including the multi-level structure of Cesium}\label{Sec:MultiLevel}

\begin{figure}
center
\includegraphics[width=\columnwidth]{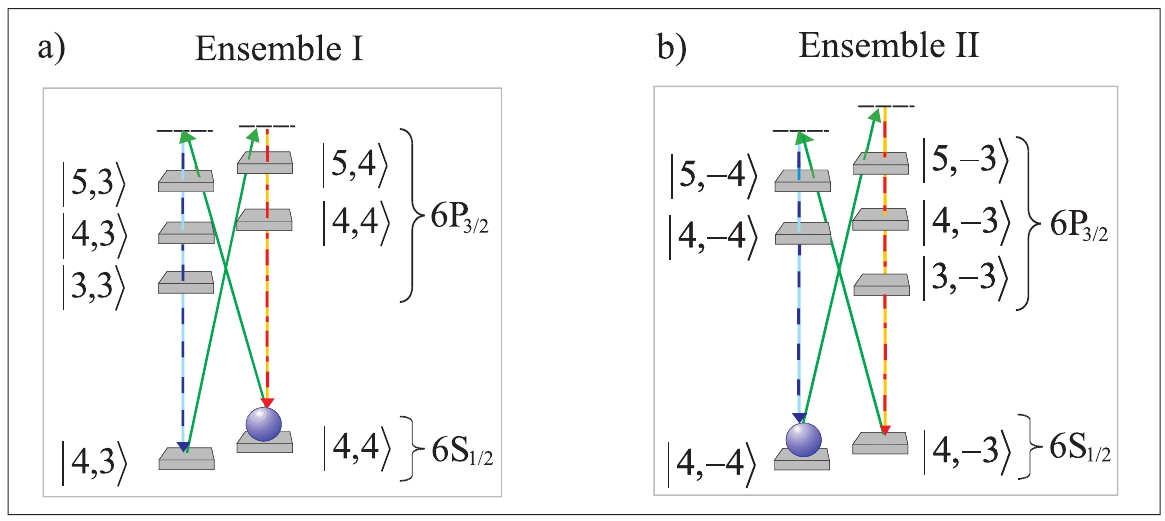}
    \caption{
    Off-resonant probing of the D$_2$ line in spin polarized $^{133}$Cs ensembles as shown in Fig.~\ref{Fig:Setup}.
    The strong coherent field in $\hat{y}$-polarization is depicted as full line, while the quantum fields corresponding to the blue and the red sideband are shown as dashed and dash-dotted lines respectively. Only desired transitions are shown - the level
    splittings are not depicted true to scale.} \label{Fig:MultiLevel}
\end{figure}

In contrast to the simple two-level model considered above,
$^{133}$Cs atoms have a multi-level structure. As illustrated in
Fig.~\ref{Fig:MultiLevel}a, a two-level subsystem can be encoded
in the $6S_{1/2}$ ground state with total spin \cite{FootnoteSpin}
$F=4$ by identifying the states $|\!\!\uparrow\rangle$ and
$|\!\!\downarrow\rangle$ with the outermost levels corresponding
to the magnetic quantum numbers (the projection of the spin along
$\hat{x}$) $m_F=4$ and $m_F=3$, such that
$|\!\uparrow\rangle\equiv|F=4,m_F=4\rangle \equiv|4,4\rangle$ and
$|\!\downarrow\rangle\equiv|F=4,m_F=3\rangle\equiv|4,3\rangle$. In
the following, we consider strongly polarized ensembles where all
atoms have been initialized in state $|\!\!\uparrow\rangle$. We
assume that only a small fraction of is transferred to state
$|\!\!\downarrow\rangle$ during the interaction and that the
population in all other levels can be neglected. The strong laser
field in $\hat{\mathbf{y}}$-polarization probes the $D_2$ line and
couples these levels off-resonantly to the excited states in
$6P_{3/2}$. This way, the passive part of the interaction
(corresponding to the transfer of atoms from
$|\!\!\downarrow\rangle$ to $|\!\!\uparrow\rangle$ and the
creation of a photon in the red sideband) involves the upper
levels with $F=4,5$, while the active part of the light-matter
interaction (corresponding to the transfer
$|\!\uparrow\rangle\rightarrow|\!\downarrow\rangle$ and the
creation of a photon in the blue sideband) involves the manifolds
with $F=3,4,5$. The corresponding $Z$-parameter can be easily
determined if the detuning and the corresponding Clebsch-Gordan
coefficients are known. In this specific case, one obtains
$Z=\left(\mu+\nu\right)=2.5$ for blue detuning $\Delta=850$MHz
with respect to the state with total spin $F=5$ within $6P_{3/2}$.

In general, $Z$ can be calculated as follows. The effective rate
for ground state transitions $|\!\!\uparrow\rangle \leftrightarrow
|\!\!\downarrow\rangle$ involving the excited state $|l\rangle$ is
given by $\Gamma_{|a\rangle\rightarrow|l\rangle\rightarrow
|b\rangle}=\Omega_{\text{R}}^2
|\frac{c_{al}c_{lb}}{\Delta_l+i\gamma_l}|^2 \gamma_l\approx
\Omega^2_{\text{R}} \frac{|c_{al}c_{lb}|^2}{\Delta_l^2} \gamma_l$,
where $\Omega_{\text{R}}^2$ is the Rabi frequency of the applied
laser field \cite{FootnoteAppliedField}, $c_{al}$ and $c_{lb}$ are
the Clebsch-Gordan coefficients for the transitions
$|a\rangle\rightarrow|l\rangle$ and
$|l\rangle\rightarrow|b\rangle$, $\gamma_l$ is the natural line
width of the excited state and $\Delta_l\gg \gamma_l$ was assumed.
If several excited states contribute, the different paths can
interfere and the effective rate for the off-resonant transition
$|a\rangle\rightarrow|b\rangle$ is therefore given by the sum
$\Gamma_{|a\rangle\rightarrow |b\rangle}=\Omega^2_{\text{R}}
|\sum_{l}\frac{c_{a l} c_{lb}}{\Delta_l}|^2 \gamma$, where the
line widths of the involved excited levels have been assumed to be
approximately equal. If the the ratio
$r^2=\frac{\Gamma_{|\downarrow\rangle\rightarrow|\uparrow\rangle}}{\Gamma_{|\uparrow\rangle\rightarrow|\downarrow\rangle}}=\frac{\mu^2}{\nu^2}$
is calculated, $Z^2=\frac{r+1}{r-1}$ can be determined.

If  the sign of the detuning is changed (using for example red
instead of blue detuning) the character of the interaction can be
changed from the predominantly passive to the active type. This
can also be achieved by interchanging the polarization of the
classical and the quantum field. As illustrated in
Fig.~\ref{Fig:MultiLevel}a, using a $\hat{\mathbf{x}}$-polarized
classical field (driving vertical transitions, in this picture)
and correspondingly a quantum field in
$\hat{\mathbf{y}}$-polarization (associated with diagonal
transitions \cite{FootnoteSpin}) would involve the excited levels
with $F=3,4,5$ for the passive part of the interaction and the
levels with $F=4,5$ for the active one (as opposed to the setting
discussed above, where it is the other way round). The imbalance
between the active and the passive part becomes less pronounced
for large detunings. If $\Delta$ is much larger than the hyperfine
splitting of the excited states, the interaction Hamiltonian can
be well approximated by $H^{\mathrm{QND}}$
\cite{WFJM09,Hammerer2010}.

\subsection{Input-output relations and characteristic properties of the interaction}\label{Sec:IORs}

In the following, the canonical quadratures
$x=(a+a^{\dag})/\sqrt{2}$ and $p=-i(a-a^{\dag})/\sqrt{2}$ will be
used. The atomic quadratures $x_A$ and $p_A$ can be identified
with the transverse components of the collective spin. Since we
consider strongly polarized atomic ensembles (see
Sec.~\ref{Sec:MultiLevel}), the macroscopic spin in
$\hat{\textbf{x}}$-direction can be described by a
${I}\!\!\!\!C$-number. The deviation from perfect
$\hat{\mathbf{x}}$-alignment is described by the collective spins
in $\hat{\mathbf{y}}$ and $\hat{\mathbf{z}}$ direction
$J_y=\sum_{i=1}^N \sigma_i^y$ and $J_z=\sum_{i=1}^N \sigma^i_z$,
where the operators $\sigma^i_y$ and $\sigma^i_z$ denote the $y$
and $z$ component of the i$^{\mathrm{th}}$ atom respectively such
that within the Holstein-Primakoff approximation
$x_A=J_y/\sqrt{|\langle J_x\rangle|}$ and $p_A=J_z/\sqrt{|\langle
J_x\rangle|}$. In terms of quadratures, the quadratic Hamiltonian
introduced above (\ref{Eq:Hamiltonian}) is given by
\begin{eqnarray}\label{Eq:HamiltonianQuadratures}
H_{\mathrm{int}}=\sqrt{2 \gamma_{s}}(Zp_A p_L(0)+\frac{1}{Z}x_A
x_L(0)).
\end{eqnarray}
As will become apparent in Sec.~\ref{Sec:Non-QNDinteraction} (see
Eq.~(\ref{Eq:QNDmagnetic})), $Z$ \cite{FootnoteSqueezing}
quantifies the squeezing (and corresponding anti-squeezing) of the
variances involved in the process \cite{WFJM09}, while
$\gamma_{s}$ \cite{FootnoteSwapping} characterizes the rate at
which atomic and light quadratures are swapped. In the balanced
case ($\mu=\nu$), $H_{\mathrm{int}}$ reduces to the QND
Hamiltoninan $H_{QND}\propto p_A p_L(0)$. Below, we introduce the
input-output relations describing the light-matter interaction and
highlight characteristic features of the imbalanced and the
balanced (QND) type.

\subsubsection{QND interaction}\label{SubSec:QND}

The balanced type of the interaction corresponding to the limit
$Z\rightarrow\infty$ (with $\sqrt{2\gamma_{s}}\cdot Z=\kappa$,
where $\kappa$ is constant), $H_{QND}=\kappa p_A p_L$ is referred
to as quantum-nondemolotion interaction since the $p$-quadratures
of atoms and light are conserved. The input-output relations for a
single cell in the absence of a magnetic field \cite{HaPC05} are
given by
\begin{eqnarray}\label{Eq:QNDbasic}
x_{A}^{out}&=&x_{A}^{in}+\kappa p_L^{in},\ \ \ \ \ \ p_A^{out}=p_A^{in},\nonumber\\
x_{L}^{out}&=&x_{L}^{in}+\kappa p_A^{in},\ \ \ \ \ \
p_L^{out}=p_L^{in},
\end{eqnarray}
where $x_L^{in}=\frac{1}{\sqrt{T}}\int_0^T dt \bar{x}_L(ct,0)$ and
$x_L^{out}=\frac{1}{\sqrt{T}}\int_0^T dt \bar{x}_L(ct,T)$
(analogous definitions hold for $p_L^{in}$ and $p_L^{out}$). Here,
the variable transformation $x_L(r,t)\rightarrow
x_L(ct-\xi,t)=\bar{x}_{L}(\xi,t)$ has been made. The spatial
variable $\xi=ct-z$ refers to a coordinate system which is fixed
to the propagating light pulse. $\xi=0$ refers to the front part
of the pulse which enters the ensemble first, while the rear part
which passes last corresponds to $\xi=cT$.

Shot-noise limited measurements of the collective spin by homodyne
detection of the light field require the application of magnetic
fields. In the presence of magnetic fields, atomic ground states
are Zeeman-shifted by the Larmorsplitting $\Omega$ as shown in
Fig.~\ref{Fig:MultiLevel} (here and in the following, we use
$\hbar=1$).

The scattering of a narrow-band classical field with central
frequency $\omega_c$ leads therefore to the emission of photons
into sideband modes, which are centered around $\omega_{c}\pm
\Omega$ in frequency space, as illustrated in Fig.~\ref{Fig:Setup}
which allows for noise reduced measurements on the light field
using lock-in methods (see \cite{Ju03}). In the time domain,
atomic information is mapped to $\sin(\Omega t)$ and $\cos(\Omega
t)$ modulated light modes
\begin{eqnarray*}
x_{f,cos}^{in}&=&\sqrt{\frac{2}{T}}\int_0^T dt f(t) \cos(\Omega t)
\bar{x}_{L}(ct,0),\\
x_{f,cos}^{out}&=&\sqrt{\frac{2}{T}}\int_0^T dt f(t)\cos(\Omega t)
\bar{x}_{L}(ct,T).
\end{eqnarray*}
$x_{f,cos/sin}$ and $p_{f,cos/sin}$ refer to a light mode with an
arbitrary envelope function $f(t)$, which varies slowly on the
time scale set by the Larmor frequency $\Omega$. The envelope
function $f(t)$ is normalized such that $\frac{1}{T}\int_0^T dt
f(t)^2=1$. In the limit $\Omega T\gg 1$, which is well fulfilled
under the experimental conditions considered here, sine and cosine
modulated modes are canonical and independent
$[x_{f,sin/cos},p_{f,sin/cos}]=i$,
$[x_{f,sin/cos},p_{f,cos/sin}]=0$. The input-output relations for
a single ensemble in a magnetic field involve an infinite
hierarchy of coupled backaction modes \cite{HaPC05}, whose
envelope functions are given by Legendre polynomials (the general
expressions can be found in \cite{HPC06}).
If a setup as shown in Fig.~\ref{Fig:Setup} with antiparallel
oriented spins, or equivalently, antiparallel oriented magnetic
fields, is considered, the input-output relations simplify
considerably since all photonic contributions except for the
lowest order cancel such that
\begin{eqnarray}\label{Eq:QNDmagnetic}
x_{A,cos}^{out}&=&x_{A,cos}^{in}+\kappa p_{L_0,cos}^{in},\\
p_{A,cos}^{out}&=&p_{A,cos}^{in},\nonumber\\
x_{L_0,cos}^{out}&=&x_{L_0,cos}^{in}+\kappa p_{A,cos}^{in},\nonumber\\
p_{L_0,cos}^{out}&=&p_{L_0,cos}^{in},\nonumber\\
\ \nonumber\\
x_{A,sin}^{out}&=&x_{A,sin}^{in}+\kappa p_{L_0,sin}^{in},\nonumber\\
p_{A,sin}^{out}&=&p_{A,sin}^{in},\nonumber\\
x_{L_0,sin}^{out}&=&x_{L_0,sin}^{in}+\kappa p_{A,sin}^{in},\nonumber\\
p_{L_0,sin}^{out}&=&p_{L_0,sin}^{in},\nonumber
\end{eqnarray}
where the EPR-operators $x_{A,cos}=(x_{A,I}+ x_{A,II})/\sqrt{2}$,
$p_{A,cos}=(p_{A,I}+ p_{A,II})/\sqrt{2}$ and $x_{A,sin}=-(p_{A,I}-
p_{A,II})/\sqrt{2}$, $p_{A,sin}=(x_{A,I}- x_{A,II})/\sqrt{2}$ have
been used. A comparison of Eq.~(\ref{Eq:QNDbasic}) with
Eq.~(\ref{Eq:QNDmagnetic}) shows that the the  input output
relations for two atomic ensembles which are Larmor-precessing in
opposite directions are formally equivalent to two independent
sets of input-output relations describing the simple case of a
single ensemble in the absence of a magnetic field.

This antiparallel setup has for instance been used for the
implementation of a quantum memory for light \cite{JSC04} and
entanglement generation between two ensembles \cite{JKP01}. Since
only the $p$-quadrature of each set of variables is mapped by the
interaction, the realization of a quantum memory required the
measurement of the $p$-quadrature of the light field and a
subsequent feedback operation on the atoms in order to transfer
both quadratures $x_{L_0,sin/cos}$ and $ p_{L_0,sin/cos}$. Also
the creation of entanglement based on a QND interaction requires
measurements on the light field. The light-matter interaction
itself does not create entanglement, but it allows for a
projection onto an Einstein-Podolski-Rosen (EPR) entangled state
with squeezed non-local variances
$\mathrm{var}\!\left(x_{A,sin}\right)$ and
$\mathrm{var}\!\left(x_{A,cos}\right)$  if $x_{L_0,sin }$ and $
x_{L_0, cos}$ are measured.

However, an ideal beamsplitter or squeezing interaction which
would allow for perfect mapping or for the creation of infinitely
entangled states in the limit $\kappa\rightarrow\infty$, can be
realized based on a QND interaction by means of a double-pass
scheme \cite{FiSOP05,QuantumMemory,SM06}, in which one of the two
contributions, $H_P$ or $H_A$ is cancelled by interference.

\subsubsection{Non-QND interaction}\label{Sec:Non-QNDinteraction}

In the following, we consider the general interaction described by
Eq.~(\ref{Eq:HamiltonianQuadratures}). The input-output relations
for a single cell in a magnetic field are given by
\begin{eqnarray*}
\left(%
\begin{array}{c}
  x_A^{out} \\
  p_A^{out} \\
\end{array}%
\right)&=& e^{-\gamma_{s}T}\left(%
\begin{array}{c}
 x_A^{in} \\
 p_A^{in} \\
\end{array}%
\right)+\sqrt{1-e^{-2\gamma_{s}T}}
\left(%
\begin{array}{c}
 x_{+,\mathrm{r}}^{in} \\
 p_{+,\mathrm{r}}^{in} \\
\end{array}%
\right),\\
\left(%
\begin{array}{c}
  x_{-,\mathrm{r}}^{out} \\
  p_{-,\mathrm{r}}^{out}\\
\end{array}%
\right)&=&e^{-\gamma_s T}\left(%
\begin{array}{c}
 x_{+,\mathrm{r}}^{in} \\
 p_{+,\mathrm{r}}^{in} \\
\end{array}%
\right)-\sqrt{1-e^{-\gamma_s T}}\left(%
\begin{array}{c}
  x_{A}^{in}\\
  p_{A}^{in} \\
\end{array}%
\right),
\end{eqnarray*}
where the exponentially rising/falling reading modes with
quadratures $x_{\pm,\mathrm{r}}$, $p_{\pm,\mathrm{r}}$ are given
by
\begin{eqnarray*}
\left(%
\begin{array}{c}
 \!\!x_{\pm,\mathrm{r}}^{in}\!\!\\
\!\! p_{\pm,\mathrm{r}}^{in}\!\!\\
\end{array}%
\right)\!\!&=&\!\!\frac{1}{2}\left(\!\left(Z\!+\!\frac{1}{Z}\right)\left(%
\begin{array}{c}
  \!\!x_{\pm,us}^{in}\!\!\\
  \!\!p_{\pm,us}^{in}\!\!\\
\end{array}%
\right)\!+\!\left(Z\!-\!\frac{1}{Z}\right)\left(%
\begin{array}{c}
  \!\!p_{\pm,ls}^{in}\!\!\\
 \!\! x_{\pm,ls}^{in}\!\!\\
\end{array}%
\right)\!\right).
\end{eqnarray*}
$x_{\pm,us}$, $p_{\pm,us}$ and $x_{\pm,ls}$, $p_{\pm,ls}$ refer to
exponentially modulated modes
\begin{eqnarray*}
\left(%
\begin{array}{c}
  x_{\pm,us}^{in} \\
  p_{\pm,us}^{in}\\
\end{array}%
\right)&=&\frac{\sqrt{2\gamma_s}}{N_{\pm}}\int_{0}^{T}d\tau
e^{\pm\gamma_s \tau}R(\tau)\left(%
\begin{array}{c}
  \bar{p}_{L}(c\tau,0) \\
  -\bar{x}_{L}(c\tau,0) \\
\end{array}%
\right),\\
\left(%
\begin{array}{c}
  p_{\pm,ls}^{in} \\
  x_{\pm,ls}^{in}\\
\end{array}%
\right)&=&\frac{\sqrt{2\gamma_s}}{N_{\pm}}\int_{0}^{T}d\tau
e^{\pm\gamma_s \tau}R(\tau)\left(%
\begin{array}{c}
 \bar{p}_{L}(c\tau,0) \\
 \bar{x}_{L}(c\tau,0) \\
\end{array}%
\right),
\end{eqnarray*}
which are located at  $\omega_c\pm \Omega$ in frequency space
respectively (compare Fig.~\ref{Fig:Setup}). The subscripts $us$
and $ls$ refer accordingly to the upper and lower sideband. The
normalization constants $N_+$, $N_-$ and the rotation matrix
$R(\tau)$ are given by
\begin{eqnarray*}
N_+&=&\sqrt{e^{\frac{\kappa^2}{Z^2}}-1},\ \
N_-=\sqrt{1-e^{-\frac{\kappa^2}{Z^2}}},\\
R(\tau)&=&\left(%
\begin{array}{cc}
  \cos(\Omega\tau) & -\sin(\Omega \tau) \\
  \sin(\Omega \tau) &  \cos(\Omega\tau) \\
\end{array}%
\right).
\end{eqnarray*}
As outlined above, the setup involving two antiparallel oriented
ensembles in magnetic fields can be conveniently described in
terms of EPR modes such that two independent sets of equations are
obtained
\begin{eqnarray}\label{Eq:IORsEbD}
\left(%
\begin{array}{c}
  \!\!x_{\text{\tiny{A,sin\!\!/\!cos}}}^{out}\!\!\\
  \!\!p_{\text{\tiny{A,sin\!\!/\!cos}}}^{out}\!\! \\
\end{array}%
\right)
\!\!\!\!&=&\!\!\!e^{-\gamma_{s}T}\!\!\left(%
\begin{array}{c}
  \!\!x_{\text{\tiny{A,sin\!\!/\!cos}}}^{in}\!\!\\
  \!\!p_{\text{\tiny{A,sin\!\!/\!cos}}}^{in}\!\! \\
\end{array}%
\right)\!\!+\!\!\sqrt{\!1\!\!-\!\!e^{-2\gamma_{s}T}}\!M_Z\!\!
\left(%
\begin{array}{c}
  \!\!p_{\text{\tiny{+,sin\!\!/\!cos}}}^{out}\!\!\\
  \!\!x_{\text{\tiny{+,sin\!\!/\!cos}}}^{out}\!\! \\
\end{array}%
\!\right),\nonumber\\
\left(%
\begin{array}{c}
  \!\!x_{\text{\tiny{-,sin\!\!/\!cos}}}^{out}\!\!\\
  \!\!p_{\text{\tiny{-,sin\!\!/\!cos}}}^{out}\!\! \\
\end{array}%
\right)
\!\!\!\!&=&\!\!\!e^{-\gamma_{s}T}\!\!\left(%
\begin{array}{c}
  \!\!x_{\text{\tiny{+,sin\!\!/\!cos}}}^{in}\!\!\\
  \!\!p_{\text{\tiny{+,sin\!\!/\!cos}}}^{in}\!\! \\
\end{array}%
\right)\!\!+\!\!\sqrt{\!1\!\!-\!\!e^{-2\gamma_{s}T}}\!M_Z\!\!
\left(%
\begin{array}{c}
  \!\!p_{\text{\tiny{A,sin\!\!/\!cos}}}^{out}\!\!\\
  \!\!x_{\text{\tiny{A,sin\!\!/\!cos}}}^{out}\!\! \\
\end{array}%
\!\right)
\end{eqnarray}
where the matrix $M_Z$ is given by
\begin{eqnarray*}
M_Z=\left(%
\begin{array}{cc}
  \!Z\! & \!0\! \\
  \!0\! & \!-\!\frac{1}{Z}\! \\
\end{array}%
\right).
\end{eqnarray*}
Due to the inherent backaction of the interaction, these
input-output relations display an exponential scaling in the
coupling strength, as opposed to Eq.~(\ref{Eq:QNDmagnetic}). This
is due to the fact that the light field is continuously mapped to
both atomic quadratures $x_A$ and $p_A$ which in turn are mapped
to the passing photonic field. This way, the light field passing
the ensembles at time $t=t_1$ is subject to an interaction which
involves the photonic contributions which have been mapped to the
atomic state during the time $t<t_1$ and experiences therefore an
effective backaction mediated by the atoms.\\

An imbalanced quadratic interaction ($|\mu|\neq|\nu|$) allows for
the realization of protocols which are not possible employing an
interaction of QND-type, for example the creation of entanglement
by dissipation which has been recently demonstrated using atomic
ensembles at room temperature \cite{EbDExperiment,EbDtheory} (see
Sec.~\ref{Sec:EntanglementByDissipation}). More specifically, the
atomic system interacts with the continuum of electromagnetic
modes $a_{\mathbf{k}}$. In the ideal case, the interaction between
the two ensembles constituting the system and the continuum of
light modes, which acts as environment, is engineered such that
the atomic system is driven into an entangled steady state. In
contrast to standard approaches \cite{DLCZ01,K08,Hammerer2010}
this method creates unconditional entanglement, since no
measurements on the light field (bath) are required
\cite{FootnoteBath}. This feature is due to the fact that the
light field possesses an infinite number of degrees of freedom,
such that a non-unitary dynamics which drives the system towards a
fixed state can be implemented. Due to this property, the
corresponding interaction is referred to as dissipative process.
Since dissipative processes are most naturally described in terms
of master equations, we will use this formalism in the remainder
of this section rather than input-output relations. Both
descriptions are equivalent. The master equation for the atomic
system discussed below can for example be obtained by considering
the interaction of atoms and light as discussed above for small
time steps $\delta t$ and tracing out the light field. The
Hamiltonian governing the light-matter interaction for two
ensembles in a magnetic field as shown in Fig.~\ref{Fig:Setup} can
be written in the form \cite{Footnote6}
\begin{eqnarray*}
H\propto \int_{\Delta\omega_{\text{ls}}}d\mathbf{k}\left(A
a_\mathbf{k}^{\dag}+A^{\dag}a_\mathbf{k}\right)+\int_{\Delta
\omega_{\text{us}}}d\mathbf{k}\left(B
a_\mathbf{k}^{\dag}+B^{\dag}a_\mathbf{k}\right),
\end{eqnarray*}
where $a_{\mathbf{k}}$ is the creation operator for a photon with
wave vector $\mathbf{k}$ and the integrals cover narrow bandwidths
$\Delta \omega_{\text{ls}}$ and $\Delta \omega_{\text{us}}$
centered around the lower and upper sideband respectively. The
atomic operators $A$ and $B$ \cite{FootnoteOperatorsDefinition}
are given by
\begin{eqnarray}\label{Eq:ABoperators}
A&=&\mu J^-_{\text{I}}-\nu J^-_{\text{II}},\\
B&=&\mu J^+_{\text{II}}- \nu J^+_{\text{I}},\nonumber
\end{eqnarray}
where $J^{+}=J_y- i J_z$ and $J^{-}=J_y + i J_z$. We assume Markov
dynamics, which is well justified for optical frequencies such
that a master equation of Lindblad form is obtained after tracing
out the photonic modes
\begin{eqnarray*}\label{masterequation}
\frac{d}{dt}\rho&=&d\ \frac{\Gamma}{2}\left(A\rho A^{\dag}
-A^{\dag} A \rho+B\rho B^{\dag} -B^{\dag} B
\rho+H.C.\right)\nonumber\\
&+&{\cal L}_{\rm noise}\rho,
\end{eqnarray*}
where $\rho$ is the reduced atomic density operator, $d$ is the
resonant optical depth of one ensemble and $\Gamma$ is the
effective single particle decay rate. The first term on the right
describes the ideal case, while the second one accounts for noise
processes. The master equation can also be derived starting from
the input-output relations (\ref{Eq:IORsEbD}) introduced above by
identifying $\gamma_s=d\Gamma$ \cite{Footnote7}. In the ideal case
(${\cal L}_{\rm noise}\rho=0$), the steady state of the
dissipative evolution is given by
$\rho_{\mathrm{EPR}}=|\Psi_{\mathrm{EPR}}\rangle\langle\Psi_{\mathrm{EPR}}|$
with
\begin{eqnarray*}
A|\Psi_{\mathrm{EPR}}\rangle=B|\Psi_{\mathrm{EPR}}\rangle=0.
\end{eqnarray*}
Since the jump operators $A$ and $B$ are nonlocal (see
Eq.~(\ref{Eq:ABoperators})), the steady state
$|\Psi_{\mathrm{EPR}}\rangle$ corresponds to an EPR-entangled
state where the collective spins in $\hat{y}$ and $\hat{z}$
direction are strongly correlated, such that
$\mathrm{var}(J_{yI}-J_{yII})+\mathrm{var}(J_{zI}-J_{zII})<|\langle
J_{xI}\rangle|+|\langle J_{xII}\rangle|$ \cite{Footnote8}.

It can be shown that this steady state is unique for $|\mu| \neq
|\nu|$ \cite{EbDtheory} (in the QND case no unique steady state
exists). This way, the desired state is reached independently of
the initial state. The initialization of the system in a well
defined fiducial state, which is typically considered a critical
issue \cite{DiVincenzo2000}, is therefore rendered unnecessary.
Moreover, the resulting state is stabilized by the dissipative
dynamics and can be maintained, in principle, for arbitrary long
times. Using these ideas, it is therefore possible to overcome
important restrictions set by the limited coherence times of
quantum systems.

\section{Experiments based on Faraday Interaction beyond QND}\label{Sec:Experiments}
In this section a series of experiments based on the described
theory are presented. In all realizations considered here, two
ensembles of Cesium atoms at room temperature are coupled to light
in a controlled fashion. This setup proves to be a versatile tool
to realize many different experiments on the quantum level
\cite{WFJM09}, including quantum communication protocols
\cite{JWK10} as well as metrology on the quantum sensitivity level
\cite{WJKRBP10}.\par
\begin{figure}
\center
\includegraphics[width=\columnwidth]{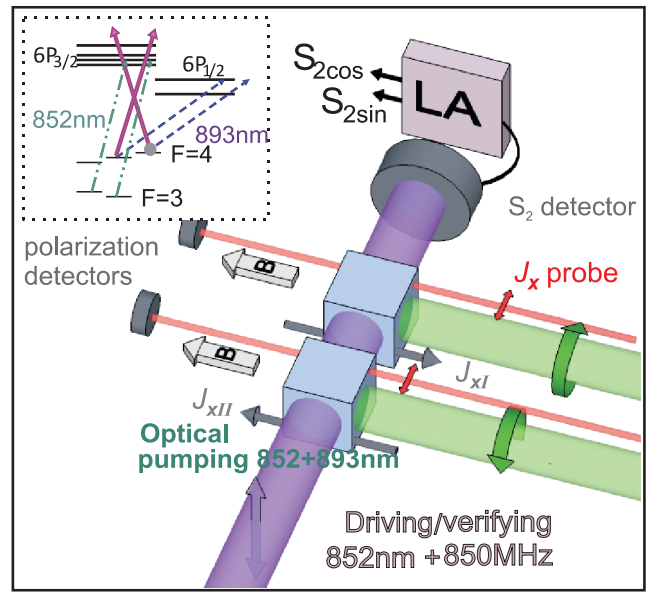}
    \caption[]{Experimental setup. The inset shows the atomic level scheme in $\hat{\mathbf{x}}$-quantization. The relevant laser frequencies and polarizations are indicated.
The dc polarization detectors measure the Faraday rotation angle
$\Theta$ proportional to the macroscopic spin $J_x$. The $S_2$
detector signal processed by the lock-in amplifier (LA). The
evaluated modes $x_{f,\text{cos}}\propto\int_0^T\cos(\Omega
t)f(t)S_2(t)dt$ and $x_{f,\text{sin}}\propto\int_0^T\sin(\Omega
t)f(t)S_2(t)dt$ with the interaction duration $T$ are used to
determine the atomic quantum spin components $J_{y,z}$ in the
rotating frame. The used mode functions $f(t)$ are mostly
exponentially decaying or growing functions. The inset shows the
optical pumping scheme.}\label{fig:set}
\end{figure}

 The basic setup is sketched in
Fig.~\ref{fig:set}. The two ensembles are prepared in oppositely
oriented coherent spin states (CSS). This is achieved by optically
pumping the atoms of the ensembles in $m_F=\pm4$ in the
$\hat{\mathbf{x}}$-direction respectively. The atoms are situated
in a magnetic field $B$ which leads to a splitting of the magnetic
sublevels by $\Omega$. The circularly polarized pump lasers are
depicted in green and the inset of Fig.~\ref{fig:set} shows the
atomic level structure, indicating laser frequencies and
polarization. The strong probe beam which is initially polarized
in $\hat{\mathbf{y}}$-direction transverses the atoms in the
$\hat{\mathbf{z}}$-direction. Behind the cells the detection
system is set up. The light observable of interest is the Stokes
operator $S_2\propto x_L$ which can be measured with polarization
homodyning techniques. The signal from the detectors is analyzed
at the Larmor frequency $\Omega$ since we are interested in the
spins  in the rotating frame. Additionally the measurement outcome
can be weighted with suitable mode-functions $f(t)$ to achieve an
optimal signal. In the following three different experiments
realized in the setup are described.

\subsection{Entanglement generated by dissipation and steady state entanglement of two macroscopic
objects}\label{Sec:EntanglementByDissipation}

The input-output-relations (\ref{Eq:IORsEbD}) arising from the
non-QND model for the scenario sketched in Fig.~\ref{fig:set}
reveal interesting possibilities, when evaluated for long
interaction times:
  \begin{eqnarray}
    x_{-,\text{cos/sin}}^{\text{out}}\!\!&\rightarrow&\!\!Z p_{A,\text{cos/sin}}^{\text{in}}, \hspace{0.3cm} p_{-,\text{cos/sin}}^{\text{out}}\!\!\rightarrow\!\! -\frac{1}{Z} x_{A,\text{cos/sin}}^{\text{in}}\nonumber\\
     x_{A,\text{cos/sin}}^{\text{out}}\!\!&\rightarrow&\!\!Zp_{+,\text{cos/sin}}^{\text{in}}, \hspace{0.3cm} p_{A,\text{cos/sin}}\!\!\rightarrow\!\! -\frac{1}{Z} x_{+,\text{cos/sin}}^{\text{in}}.\label{Limit}
    \end{eqnarray}
The two systems, light and atoms, swap state and individually get
squeezed by the factor $Z^2$ if both systems start in a minimum
uncertainty state. In \cite{WFJM09} the observation of two mode
squeezing of light in $p_{-,\text{cos}}^\text{out}$ and
$p_{-,\text{sin}}^\text{out}$ is reported. There a noise reduction
of 3dB was achieved. At the same time the equations predict a
reduction in the noise of the atomic operators $p_{A,\text{cos}}$
and $p_{A,\text{sin}}$, indicating a possibility to achieve
entanglement between the two ensembles via this light-atom
interaction. Atomic entanglement is of special interest as it can
in principle be distributed and then stored until one wishes to
use it. However, exactly the storage represents a major problem in
most previously conducted atomic entanglement experiments. Atomic
entangled states proved to be extremely fragile, whether they were
generated by mapping of squeezed light onto atoms
\cite{Honda2008,Appel2008}, by measurement
\cite{JKP01,ChRFPEK05,EiAMFZL05,MCJ06,YCZC08,AWO09}, atomic
interactions \cite{GZN10,RB10} or a nonlinear interaction mediated
by light \cite{FKJ08}. The coupling to the environment leads to
decoherence which until now irresistibly was followed by the
disappearance of entanglement after a certain time. Several
proposals have been made, as to how to use carefully engineered
environments to create a situation where entanglement is reached
by dissipation to overcome this shortcoming
\cite{PH02,KC04,DMK08,VWC09,BMS11}. More specifically,
dissipatively generated entanglement was proposed for our system
in \cite{EbDtheory}. This theoretical approach is an extension of
the presented theory, in which spontaneous emission in the
continuum of modes and atomic decay mechanisms are considered. The understanding gained this way is that entanglement between the two ensembles is generated by the interference of different processes in the two ensembles for which an indistinguishable photon is emitted into the common mode. The processes in the forward direction are collectively enhanced and a photon emitted into for example the upper sideband stands for an atomic excitation in ensemble one, or an annihilation in ensemble two.\\

\begin{figure}
\center
\includegraphics[width=\columnwidth]{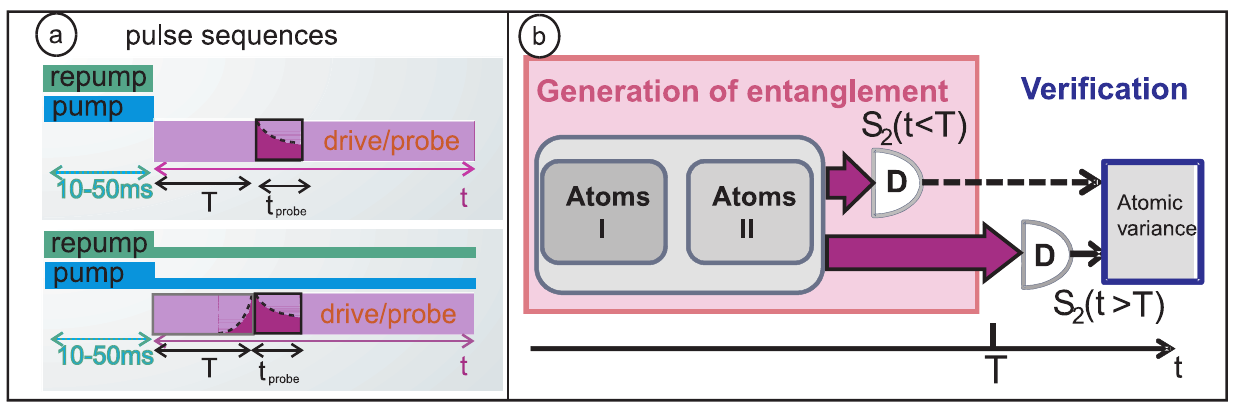}
    \caption{(a) Pulse sequence. In the upper sequence a CSS is prepared by optical pumping after which the atomic state evolves in the presence of the probe light. Below a pulse sequence for the experimental realization of a steady state scenario with high atomic population of the relevant atomic sublevels is shown. Here, the pump lasers are not completely turned off but ramped down to optimized strength.(b) Schematic illustration of entanglement generation in the steady state and verification. The signal taken at $t<T$ is given to the verifier as additional
information to reduce the noise of the collective atomic
operators. The signal from the detector D for times $t>T$ is used
for verification of entanglement. This corresponds to the pulse
sequence presented at the bottom of (a).}\label{fig:dis}
\end{figure}

The pulse sequence for an entanglement generation experiment is
shown on top in Fig.~\ref{fig:dis}a. However, when the experiment
was carried through, the proposed unconditional entanglement
generation procedure lead only to a long entanglement duration of
around 15ms, but not the wanted steady state entanglement. The
reason is the loss of atoms from the atomic two level system of
relevance due to decay. To counteract this depopulation, two pump
lasers are added. First a pump laser on resonance with the
$F=4\rightarrow F'=4$ transition on the $D_1$ line is added
(depicted by the blue lines in the inset of Fig.~\ref{fig:set}),
for which $m=4$ is a dark state. This incoherent process leads to
an increase in the duration of unconditional entanglement to $\sim
40ms$. In this scenario atoms still undergo transitions to the
$F=3$ ground state, effectively reducing the number of atoms
participating in the interaction. To avoid this depletion, a
repump beam is added (green lines in the inset of
Fig.~\ref{fig:set}). In this experimental setting (bottom of
Fig.~\ref{fig:dis}a), a steady state is achieved after few ms, but
no entanglement can be deduced from the noise of the collective
atomic operators. However, when a measurement on the light output
is added to the protocol \cite{JKP01}, a steady state which is
entangled conditioned on the continuous measurement outcomes of
$x_L$ arises. The principle of this procedure is sketched in
Fig.~\ref{fig:dis}b. In \cite{Krauter2011} such a long time
entanglement was measured for up to an hour, where a 1dB noise
reduction was achieved.

\subsection{Quantum memory for entangled two-mode squeezed states}

A quantum memory for light is a key element for the realization of
future quantum information networks. The basic principle of such a
memory protocol is the transfer of the quantum state of light to a
storage medium. Here, this means the canonical quantum variables
of light are transferred to the corresponding atomic ones. We know
from the input-output equations presented in section
\ref{Sec:IORs}, that if we could achieve the long interaction time
regime, the underlying interaction would swap the states of the
two systems (see Eq.~(\ref{Limit})). However, as this is not
possible for our experimental realization due to decoherence, a
trick is applied to achieve the desired state transfer; the
$x_{L_0,\text{cos/sin}}$ of the outgoing light are measured and
the measured results are fed back to the atoms via RF magnetic
fields: $p_{A,\text{cos/sin}}^\text{out}-g\cdot
x_{+,\text{cos/sin}}^\text{out}$. Following eq. (\ref{Eq:IORsEbD})
and assuming a coupling constant $\kappa=\sqrt{1-e^{-2\gamma_s
T}}\cdot Z=1$ and a feedback gain $g$ which is also $1$, the
collective atomic operators are then left as:
 \begin{eqnarray}
 P_{A,\text{cos/sin}}^\text{fin}&=&-x_{+,\text{cos/sin}}^{\text{in}},\\
 x_{A,\text{cos/sin}}^\text{fin}&=&\sqrt{1-\frac{1}{Z^2}}
 x_{A,\text{cos/sin}}^{\text{in}}+p_{+,\text{cos/sin}}^{\text{in}}\nonumber
 \end{eqnarray}
\begin{figure}
\center
\includegraphics[width=\columnwidth]{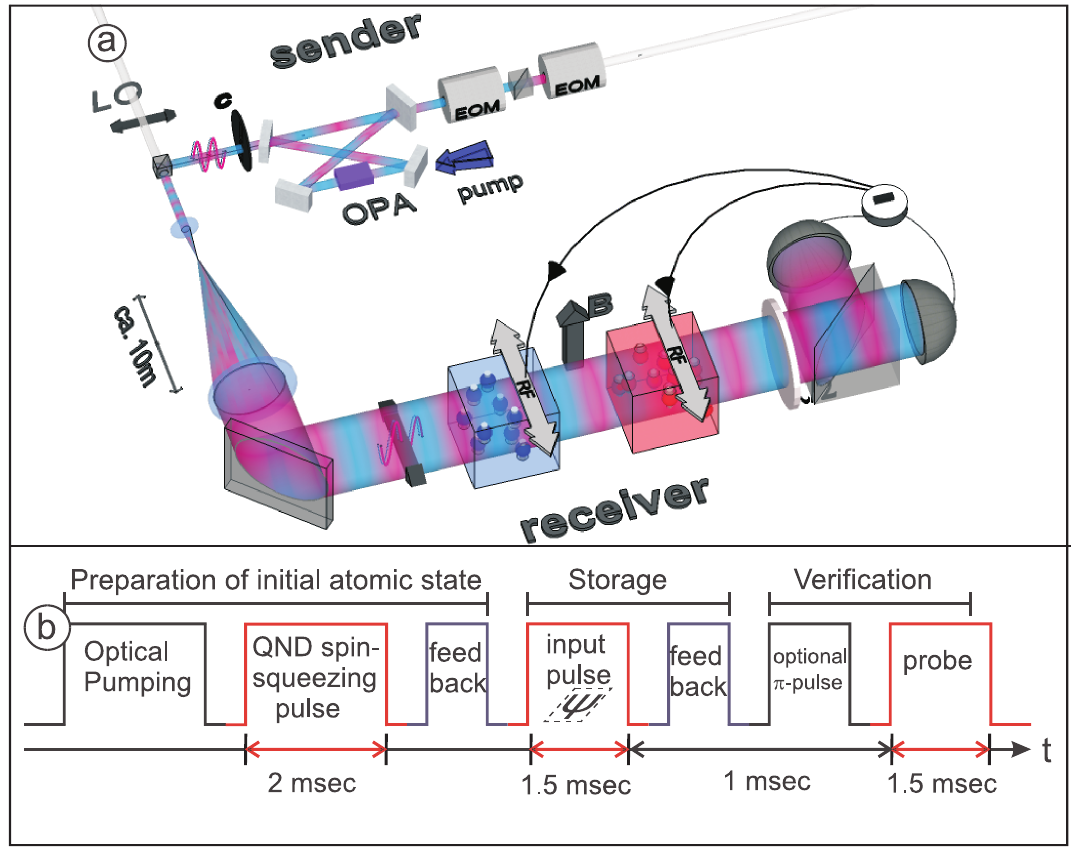}
\caption{Two-mode entangled (squeezed) light is generated by an
optical parametric amplifier (OPA). A variable displacement of the
state is achieved by injecting a coherent input into the OPA which
is displaced with help of electro-optical modulators (EOM). The
output of the OPA is shaped by a chopper, and combined on a
polarizing beamsplitter with the local oscillator (LO) beam, such
that the squeezed light is only on during the second probe pulse.
A beam shaper and a telescope create an expanded flat-top
intensity profile. The light is then send to the memory consisting
of two oppositely oriented ensembles and the homodyne detection
system. The detector signal is processed electronically and used
as feedback onto the spins via RF magnetic field pulses. Below,
the pulse sequence is shown. After preparing the squeezed spins
state, the actual storage takes place, followed by a verification
pulse.}\label{Fig:Mem}
\end{figure}
Clearly the input light state is mapped onto the atoms with some
additional noise coming from the atomic $x$ operators. Compared to
a protocol based on a QND interaction and feedback, this
additional noise is suppressed.
 A memory
based on a similar protocol was conducted for coherent states in
the setup exceeding the achievable fidelity for any classical
memory \cite{JSC04}. The next obvious step was to map non
classical states, like two mode squeezed light states, in other
words states possessing Einstein-Podolsky-Rosen (EPR)
entanglement. A squeezed light source \cite{SSP02} was used to
produce displaced squeezed states, which were mapped onto the
atoms. The setup is shown in Fig.~\ref{Fig:Mem} and described
briefly in the caption and in more detail  in \cite{JWK10}.\par
The experiment was refined by employing an additional probe pulse
after the preparation of the atomic CSS to reduce the input noise
of $x_{A,\text{cos/sin}}$. The achieved squeezing was
approximately $-14\%$. The initial light state was squeezed by
6dB.
\par To evaluate the performance of the memory the fidelity of certain sets
of input states was calculated and compared to a classical
benchmark also presented in \cite{JWK10}. The classical benchmark
was surpassed for a certain input set with a square displacement
range with a maximum displacement of 3.8 and two possible
squeezing phases.

\subsection{Quantum noise limited and entanglement-assisted magnetometry}
\begin{figure}
\center
\includegraphics[width=\columnwidth]{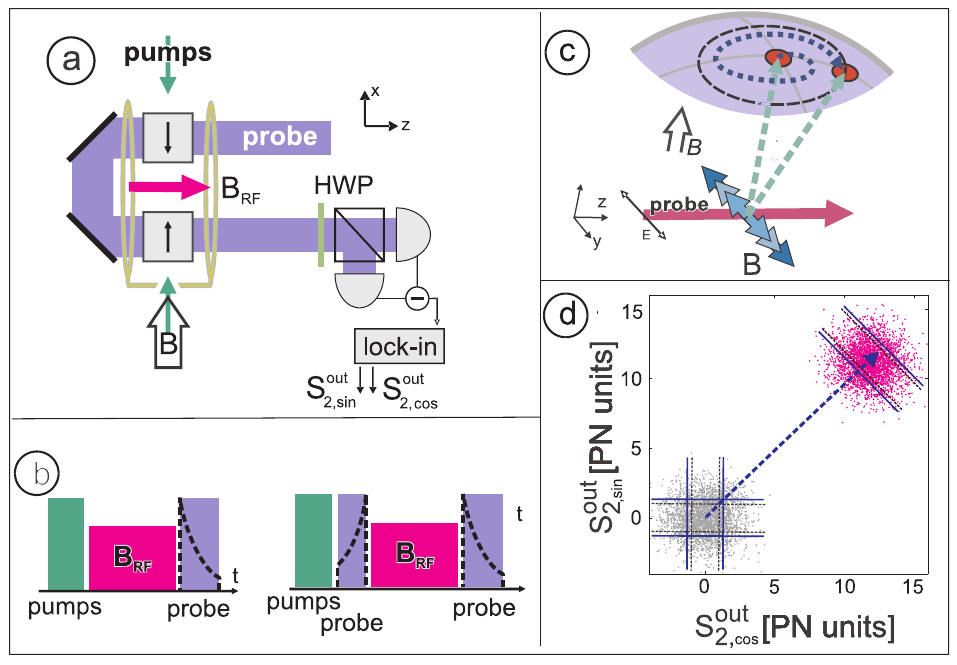}
    \caption[]{(a) The experimental setup is similar to the usual settings. A pulse of $B_\text{RF}$ at
the frequency $\Omega$ is applied orthogonally to the B field,
such that the created displacement in the spin lies in the same
direction for both cells. (b) The pulse sequence for projection
noise limited magnetometry is also similar to previous
experiments. The temporal mode function for the probe is indicated
as the dashed black curve. The pulse sequence to the right shows
the scenario including the temporal modes for
entanglement-assisted magnetometry. (c) Function principle of
radio-frequency magnetometer. The atomic spin J precesses in
crossed dc and rf magnetic fields (the blue dashed spiral). The
state of the spin after the rf pulse (black dashed circle) is
measured using a probe beam, on which the precessing J imposes
oscillating polarization rotation. (d) Experimental results. Grey
points are the experimental data for for the spin projection on
y¡z plane in the rotating frame for a series of measurements of
the polarization rotation signal $S_2^\text{out}(t)$ before the
$B_\text{RF}$ is applied. The blue and the dashed black lines show
the standard deviation of the total noise, and the contribution
due to the projection noise (PN), respectively. The spin
precession as illustrated in (c) is in corresponds to the
displacement in the rotating frame indicated by the dashed blue
arrow in (d). Results of the series of measurements taken with a
$B_\text{RF}$ = 36 fT rf field applied over 15 ms are shown as
pink points. The rf field is calibrated using a pick-up coil. When
the entangling probe pulse is applied as explained in the text,
the grey points and the pink points become correlated which leads
to a reduced spin noise and improved sensitivity.}\label{fig:magn}
\end{figure}

Oriented atomic ensembles can be used as a sensor for magnetic
fields. The realization of quantum noise limited experiments in
the presented setup in the past \cite{JSC04,SKO06} laid the basis
for a high performing atomic magnetometer presented in
\cite{WJKRBP10}. Ultra-sensitive atomic magnetometry is usually
based on the measurement of the polarization rotation of light
transmitted through an ensemble of atoms placed in the magnetic
field \cite{Budker2007}. For $N_A$ atoms, the magnetic moment
(spin) of the optically pumped ensemble has the length
$J_{x}=4N_A$. A magnetic field along the $y$ axis causes a
rotation of the spin in the $x-z$ plane. The corresponding
displacement of the transversal spin $J_y$ will be proportional to
the strength of the applied magnetic field and also to the
macroscopic spin $J_x$. Also, the longer the exposure duration
$\tau$ to a given magnetic field, the bigger the caused rotation.
However, the decoherence time $T_2$ of the transversal spin sets a
limit to the optimal duration $\tau$. The introduced light atom
interface can now be utilized to read out the caused spin
rotation. Polarization of light propagating in $z$-direction will
be changed due to $J_z$ (similar to the Faraday effect), as can be
seen from the input-output equations where $x_L\propto S_2$ is
changed according to $p_A\propto J_z$. This measurement is limited
by quantum fluctuations (shot noise) of light and the projection
noise (PN) of atoms. Quantum back-action noise of light onto atoms
is avoided by the antiparallel initialization  of the two
ensembles \cite{JKP01}. As shown rigorously in \cite{Tsang2011},
the backaction cancellation method applied here is the most
general way of measuring ac fields and forces with the sensitivity
beyond the Standard Quantum Limit (SQL) which leads to achieving
the Quantum Cramer-Rao bound of sensitivity. PN originates from
the Heisenberg uncertainty relation $\mathrm{var} (J_z)
\cdot\mathrm{var} (J_y) \geq J_x^2/4$, and corresponds to the
minimal transverse spin noise $\delta J_{z,y} =\sqrt{2N_A}$ for
uncorrelated atoms in a CSS \cite{WBI92} where $\delta J_{z,y}$ is
referring to the standard deviation. Here we are looking at atomic
ensembles in a bias magnetic field with $B\approx0.9G$ which
causes the atomic spins in y- and $\hat{\mathbf{z}}$-direction to
precess at the Larmor frequency $\Omega=2\pi\,322$kHz. A magnetic
RF-field with frequency $\Omega$ causes a displacement of the
atomic spin as illustrated in fig. \ref{fig:magn}c.
To optimize the decoherence time $T_2$, while the RF field is
turned on, all laser fields are turned off. The pulse sequence of
relevance is shown on the left of fig.\ref{fig:magn}b. After the
RF field, the displacement is read out via $S_2$ which is analyzed
at the frequency $\Omega$. The measurements are weighted with
suitable exponentially decaying modefunctions which give the best
signal to noise ratio (SNR). In fig.\ref{fig:magn}d a scatter plot
of measurement outcomes for a specific realization is shown. In
another setting, for $N_A=1.5\cdot10^{12}$ and $\tau=22ms$ a
sensitivity of $4.2(8)\cdot 10^{-16}$Tesla$/\sqrt{Hz}$ was
achieved approaching the best to-date atomic rf magnetometry
sensitivity \cite{lss06} obtained with $10^{4}$ times more atoms.

 The achieved performance lies around 30$\%$ above the PN limit. The residual noise sources arise due to the decay of the spin and from the SN of light - which is suppressed due to the "non-QND" type of interaction. \par

In earlier works \cite{JKP01,Quint} it was shown that entanglement
between two atomic ensembles can be generated via a measurement on
light that has interacted with both ensembles \cite{JKP01}. In
principle it should be possible to improve the sensitivity by
venturing away from the CSS and towards such two mode squeezed
atomic states. The drawback is that due to the short lifetimes of
the squeezing compared to the optimal exposure time, the optimal
setting cannot be improved in such a way. However, it is possible
to improve the measurement performance for shorter RF-pulses or
larger bandwidths.  An additional probe pulse was used to
conditionally squeeze the atomic input operators prior to the
exposure to the RF field. In \cite{WJKRBP10} it was shown that an
increase in the SNR can be seen for short pulses when an
entangling step (see rigth pulse sequence in Fig.~\ref{fig:magn})
was added.
\subsection{Outlook}
One future perspective of the presented setup lies in engineering
miniaturized gas cells with a cross section of $200\times200 \mu
$m$^2$ opening up for the possibility of smaller magnetic field
sensors as well as a small fiber integrated cell network. Moving
to new setup designs also gives the opportunity to decrease the
effect of the main limiting factor of all presented experiments:
decoherence. Here, decoherence arises amongst others from
collisions with the wall, magnetic field instabilities and
spontaneous emission. The resulting decay of the spin, reduces the
achievable degree of entanglement, the mappping-fidelity and  the
sensitivity of magnetic field measurements. To diminish the effect
of spontaneous emission, one approach could be the inclusion of a
bad cavity around the next generation of micro cells. This
enhances the collective effect on the atoms which lays the basis
for all presented experiments without increasing the spontaneous
emission. The effect of wall collisions can presumably be
decreased by working with recently developed coatings
\cite{BKLB2010}. In alkali-metal vapor cells prepared with such
coatings lifetimes of the spin up to one minute have been
observed.

\section{Heisenberg scaling in entanglement assisted atomic
metrology}

Atoms of an ensemble in a spin squeezed state (SSS) are entangled
~\cite{SoDCZ00} if $ (\delta J_z)^2<\frac{|\langle
J\rangle|^2}{N_A}\Rightarrow \xi\equiv\frac{(\delta
J_z)^2}{|\langle J\rangle|^2}N_A<1$ where $ J_z$ is one of the
collective (quasi)-spin components orthogonal to the mean spin
direction and $\xi$ defines the squeezing parameter. Under this
condition the state also improves the signal-to-noise ratio in
atom interferometry, metrology and sensing ~\cite{WINELAND1992}.

Generation of such SSS fulfilling the above condition in an
ensemble of $\sim10^5$ atoms via a QND measurement of $J_{z}$. was
reported in ~\cite{AWO09}. The quasi-spin corresponded to the two
clock levels of Cs atoms. Later it was shown that this SSS
improves the precision of an atomic clock ~\cite{clock2010}. As
discussed in ~\cite{Saffman2008}, the degree of spin squeezing
scales with the optical depth $d=\sigma_0 N_A/\mathcal{A}$ (with
scattering cross section on resonance $\sigma_0$, number of atoms
$N_A$, and beam cross section $\mathcal{A}$) as $\xi=(1/(1+d\eta )
+a \eta)/(1- \eta)^2$ where $\eta$ is the probability of
spontaneous emission caused by the QND measurement. The first term
in paranthesis describes the $J_z$ noise reduction due to the QND
measurement, the second term describes the change in the $J_z$
component due to the spontaneous emission and the factor $(1-
\eta)$ is responsible for the shortening of the macroscopic spin
due to the spontaneous emission. The constant $a$ depends on the
particular level scheme and details of the QND interaction.

\begin{figure}[t!]
  \begin{center}
  \includegraphics[keepaspectratio,width=\columnwidth]{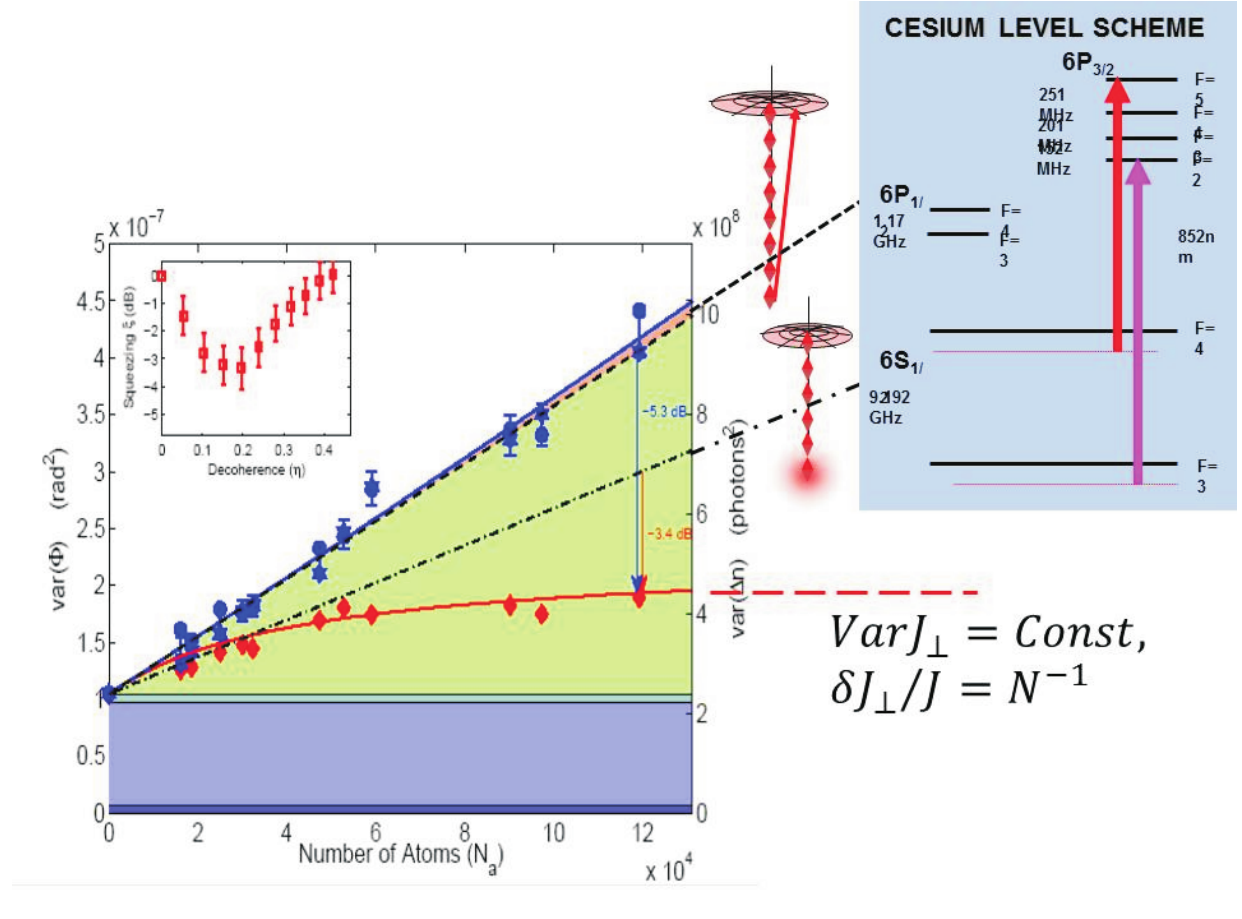}
  \end{center}
  \caption[Spin squeezing and Heisenberg scaling]{ Blue points, stars and dashed line:
    Variances $\mathrm{var}(\phi_1)$, $\mathrm{var}(\phi_2)$ of the $J_z$ spin noise
   variance of atoms in a CSS proportional to $N_A$ .  Dash-dotted line: equivalent CSS
    projection noise reduced by the loss of atomic coherence. Red
    diamonds and red line: reduced noise of SSS which asymptotically approaches a constant level independent of the number of atoms.
    Blue fields: optical shot noise (light blue) and detector noise
    (dark blue).  \textbf{Inset}, Atomic level scheme and two QND probes.}\label{Fig:Heisenberg}
\end{figure}

The QND measurement of the clock state population difference in
~\cite{AWO09} is realized by detecting the state dependent phase
shift of two off-resonant probe laser beams using a Mach-Zehnder
interferometer. One probe $P_\downarrow$ is coupled to the state
$|{\downarrow}\rangle \equiv 6 S_{1/2}(F=3,m_F=0 )$, while a
second probe $P_\uparrow$ is coupled to the state
$|{\uparrow}\rangle \equiv 6 S_{1/2}(F=4, m_F=0) $ (see
Fig.~\ref{Fig:Heisenberg}B). Cold Cs atoms are loaded into an
optical dipole trap, aligned to overlap with the probe arm of the
MZI, and a CSS $\bigotimes_{i=1}^{N_A}\left[\frac{1}{\sqrt{2}}
  \bigl(|{\downarrow}\rangle+|{\uparrow}\rangle\bigr)\right]_i$ is prepared. Successive QND measurements on
the sample are performed, after which all atoms are pumped into
the $F=4$ level to determine the total atom number $N_A$. The
sequence is repeated several thousand times for various $N_A$.

The dichomatic QND measurement with cyclic transitions does not
add any noise to $\delta J_{z}^{2}$ as elaborated
in~\cite{Saffman2008} which corresponds to $a=0$ in the expression
for $\xi$. Hence the optimal squeezing $\xi_{\min } \propto
1/\sqrt{N_A}$ is expected for $\eta =\frac{1}{3}$, assuming a
large resonant optical depth $d_0$ where $N_A$ is the number of
atoms. The precision of the determination of the macroscopic spin
direction then scales as $\delta J_z/(J) \propto\sqrt{\xi /N_A}
\propto (1/N_A)$ which is the Heisenberg scaling.

Fig.~\ref{Fig:Heisenberg} demonstrates approaching the Heisenberg
scaling. The atomic spin noise of the SSS becomes independent of
the $N_A$ for large atomic numbers, the feature that ensures
Heisenberg scaling for the precision of the spin direction $\delta
J_z/(J)$.

\section{Interface to solids}\label{Sec:Ensembles+Solids}

In this last section we will turn to yet another avenue which is
opening up for experiments with room temperature vapors of neutral
atoms, and the perspective for quantum information processing
associated to it. This avenue promises to lead to the integration
of the Faraday based light matter interface to solid state
systems, in particular to hybrid quantum systems
\cite{Wallquist2009} of atomic ensembles and micro- or
nanomechanical oscillators. The coupling of the latter systems to
light has recently become the focus of the burgeoning field of
optomechanics. Strong light-matter coupling \cite{GROEBLACHER1992}
and optical cooling of mechanical oscillators close to their
ground states \cite{GROEBLACHER1992} have been seen in recent
experiments. For recent reviews on this field see
\cite{AspelmeyerReview}.

The physics of an optomechanical system can be understood from the
simple picture of a harmonically bound, moving mirror, which
provides one end mirror of a Fabry-Perot cavity. Small
displacements of the mirror from its equilibrium position will
result in phase shifts of the cavity field, i.e. shifts of the
phase quadrature $p_c$ depending on the mirror position $x_m$
\cite{FootnoteMirror}. In turn, the radiation pressure force will
change the mirror momentum, which amounts to a change of the
mechanical momentum $p_m$ depending on the cavity's amplitude
quadrature $x_c$. In a picture where only linear effects of this
mutual changes are considered, we see that the resulting dynamics
can be described by relations similar to the ones given in
Equations~\eqref{Eq:QNDbasic} with $x_{A(L)},~p_{A(L)}$ being
replaced by $p_{c(m)},~x_{c(m)}$ respectively. In the bad cavity
limit, where the intracavity field can be adiabatically eliminated
from the dynamics this statement will just as well hold true for
the mechanical quadratures and the quadratures for a propagating
pulse being reflected off the optomechanical system. This
reasoning so far neglects the free oscillatory motion of the
harmonically bound mirror at frequency $\omega_m$. In a regime
where the pulse length $T$ is much shorter than a period
$1/\omega_m$ this is well justified. The associated QND
measurement of the mechanical displacement has recently been
discussed in detail for a current optomechanical setup in
\cite{Vanner2011}. In the other limit where $T\gg 1/\omega_m$ the
oscillatory motion has to be taken into account and the
input-output relations for the mechanical system and the
propagating pulse are in fact equivalent to the ones for an atomic
ensemble in a magnetic field causing a Larmor splitting of
$\Omega=\omega_m$ interaction with a pulse in QND fashion, as
given in Sec.~\ref{SubSec:QND}.

This analogy lies at the heart of the interface of atomic
ensembles to solids suggested in \cite{Hammerer2009}, and will be
summarized in the following. It was explained in
Sec.~\ref{SubSec:QND} how the QND interaction between an atomic
spin and light can be realized. We have also seen that placing two
atomic ensembles in magnetic fields will give rise to a QND
interaction of light with EPR-operators associated to the
transverse spin components of the two ensembles, cf.
Equ.~\eqref{Eq:QNDmagnetic}. The idea of the interface to solids
is to apply this method to a hybrid system consisting of an atomic
ensemble and an optomechanical system. While the latter
necessarily has an effective positive ``Larmor'' frequency
$\Omega=\omega_m>0$, the atomic ensemble can in turn be used to
effectively realize a mechanical oscillator of \textit{negative
mass} with Larmor frequency $\Omega=-\omega_m$. For mechanical
oscillators with typical resonance frequencies of several $100$kHz
this requires only moderate magentic field strengths. Overall,
when a sufficiently long light pulse interacts first with an
optomechanical system and then with an atomic ensemble tuned to
$\Omega=-\omega_m$ and in QND mode, then the overall input-output
relation is given by Eqns.~\eqref{Eq:QNDmagnetic}, with
$x_{A,cos}=(x_A+p_m)/\sqrt{2}$, $p_{A,cos}=(p_A-x_m)/\sqrt{2}$ and
$x_{A,sin}=(x_A-p_m)/\sqrt{2}$, $p_{A,sin}=(p_A+x_m)/\sqrt{2}$
describing now hybrid EPR operators involving the atomic spin
quadratures $x_A,~p_A$ and mechanical position and momentum
operators $x_m,p_m$. A homodyne measurement of of light will then
project the hybrid system in an entangled EPR state, which can in
principle serve as a resource for teleportation protocols.

The important point made in the original proposal
\cite{Hammerer2009} was to show that the parameters of these two
very different systems -- a nanomechanical oscillator and a
collective atomic spin -- can be matched in such a way that it
becomes possible to establish an interface in the sense described
above. Apart from matching Larmor to mechanical resonance
frequencies this requires also that the optomechanical coupling
strength can be of similar magnitude than the one of the light
atoms interface. We have seen that in the latter system it is
essentially only the parameter $\kappa$ which enters the input
output relations. In the optomechanical system the equivalent
parameter turns out to be
\[
\kappa_{OM}=2 kx_{ZPF}\sqrt{N_{ph}}\mathcal{F},
\]
where $k$ is the wave number, $x_{ZPF}=\sqrt{\hbar/2m\omega_m}$
the zero point fluctuation of the mechanical oscillator, $N_{ph}$
the number of photons and $\mathcal{F}$ the cavity finesse of the
optomechanical system. In \cite{Hammerer2009}, we demonstrated
that it is possible to have $\kappa_{OM}\simeq\kappa\simeq 1$
under compatible experimental conditions.

One major difference of the mechanical system as compared to the
atomic spin is of course that the preparation of the ground state
can be achieved very efficiently in atoms via optical pumping
while it is a much more demanding task on the side of the
mechanical oscillator. While these systems can provide very high
quality factors $Q=\omega_m/\gamma$ on the order of $10^6$ (with
$\gamma$ the width of the mechanical resonance), there is still a
rather large mean occupation $\bar{n}=k_BT_0/\hbar\omega_m$ in
thermal equilibrium at ambient temperature $T_0$, and associated
to it a comparatively large thermal decoherence rate
$\gamma\bar{n}$. It turns out that the protocol described above is
remarkably resilient to the initial thermal occupation of the
mechanical oscillator. For an initial thermal occupation $\bar{n}$
homodyne detection of $x_{L_0,cos}^{out}$ and $x_{L_0,sin}^{out}$
will prepare an EPR squeezed state with reduced EPR variance
\[
\Delta\left(p_{A,cos}^{out}\right)^2+
\Delta\left(p_{A,sin}^{out}\right)^2=
\left[\frac{1}{1+\bar{n}}+2\kappa^2\right]^{-1}.
\]
An entangled state will be produced if the right hand side falls
below one, which can be achieved for moderate values of $\kappa$
even for mean initial occupations much larger than one. This can
be understood by noting that -- for large $\kappa$ -- entanglement
is here created in a projective measurement, such that the initial
state and its entropy become irrelevant for a suffiently strong
QND measurement. The state thus created opens an EPR channel of
entanglement between optomechanics and atomic spins, a basis for
quantum state transmission or transduction
\cite{Rabl2009,Stannigel2010}.

Overall, it is remarkable to see that these very disparate systems
realize very similar physics in that the light matter interaction
is described by the same equations. Moreover, also the time scales
of the dynamics in both systems can be comparable and are
compatible for combination and interfacing. Other possible
realizations of such ideas were worked out in
\cite{Hammerer2009b},\cite{Wallquist2010} and
\cite{Hammerer2010b}. Note that micro- and nanomechanical systems
can not only be coupled to atoms, as demonstrated here, but also
to many other systems, such as e.g. spin impurities, electrical
(superconducting) circuits etc. \cite{Rabl2009,Stannigel2010}. In
the long run we thus expect that these systems can  play an
important role as transducers for quantum information in
architectures for quantum information processing.

\section{Conclusions}

We have reviewed the recent developments in the light-matter
interface based on the Faraday interaction of light with room
temperature atomic vapors. These developments are largely based on
an extension of this dynamics from the well known and established
QND interaction to a regime of more general interaction with a
tunable balance between the components of a passive beam splitter
and an active down conversion dynamics. The realization of
improved magnetometry, quantum memory for squeezed state and the
preparation of steady state entanglement by dissipative dynamics
are all based on this tunability of the Faraday interaction. We
also reviewed possible combinations and interfaces of collective
atomic spins with nano- or micromechanical oscillators, providing
a link between atomic and solid state physics approaches towards
quantum information processing.

In the past decade quantum interfaces between neutral room
temperature objects and optical photons have been extensively
explored by a number of leading groups. Quantum state transfer
between light and atoms, such as quantum memory and quantum
teleportation, entanglement of massive objects, as well as
measurements and sensing beyond standard quantum limits have been
demonstrated. One promising direction for the future developments
in this field is to develop a robust, integrated and scalable room
temperature atom-light interface and to incorporate it into a
hybrid multi-facet quantum network with other relevant quantum
systems, such as nano-mechanical oscillators and electronic
circuits. Micro-size room temperature atomic quantum memories in
spin protecting micro-cells appear to be excellent candidates for
this task. This research thus adds to the highly interdisciplinary
effort to enable large scale quantum information processing, be it
for long distance quantum communication, distributed quantum
computation or scalable photonic quantum computers.

\begin{acknowledgements}
We acknowledge support from the Elite Network of Bavaria (ENB)
project QCCC and the EU projects MALICIA and QUEVADIS. H. K.
acknowledges funding through the Centre for Quantum Engineering
and Space-Time Research (QUEST) at the Leibniz University Hanover.
\end{acknowledgements}

%
%
%

%

\begin{thebibliography}{10}
\bibitem{FootnoteSpatiallyLocalizedModes}
$a_{L}^{\dag}(r)=\frac{1}{\sqrt{2\pi}}\int_b d\omega \
a^{\dag}_{\omega}e^{-i \left(\omega-\omega_{\text{L}}\right)
\frac{r}{c}}$, where $a^{\dag}_{\omega}$ is the creation operator
for a photon with wave vector $k$, $c$ is the speed of light, and
$b$ and $\omega_{\text{L}}$ are the bandwidth and central
frequency of the applied laser field.
$[x_{L}(z),p_{L}(z')]=ic\delta_b(z-z')$, where the deltafunction
has a width of the order $c/b$
\cite{SiD03,MaM04,HaPC05,Hammerer2010}.
\bibitem{FootnoteGaussian}
Here and in the following, Gaussian input states and Gaussian
measurements (homodyne detection on the quadratures of the light
field) are considered. The time evolution according to a quadratic
Hamiltonian is a linear transformation which preserves the
Gaussian character of the involved quantum states. Therefore, the
setting under consideration can be conveniently described using
the Gaussian formalism.
\bibitem{FootnotePointlike} We assume here
a pointlike atomic ensemble located at $z=0$.
\bibitem{FootnoteSpin} The quantization axis is defined by the
direction of the magnetic field. Therefore, we use here
quantization along $\hat{\mathbf{x}}$.
\bibitem{FootnoteAppliedField} $\Omega_{\text{R}}^2$ is
proportional to the intensity of the applied laser light and
independent of the excited state of the atomic transition driven
by this field.
\bibitem{FootnoteSqueezing}
$Z=\sqrt{\frac{a_1}{14a_2}}$, where $a_1$ and $a_2$ are the atomic
vector and tensor polarizabilities \cite{WFJM09}. The calculation
of these values can be found for example in \cite{Hammerer2010}.
\bibitem{FootnoteSwapping}
$\gamma_s=14 a_1 a_2 \frac{\Phi N_A}{A^2}\left(\frac{\gamma
\lambda^2}{8 \Delta 2\pi}\right)^2$, where $\gamma$ is the natural
line width, $\Phi$ is the photon flux per unit time, $A$ is the
cross section of the applied light beam, $\lambda$ is the wave
length and $\Delta$ is the detuning.
\bibitem{FootnoteBath}
Note that standard protocols such as the DLCZ scheme \cite{DLCZ01}
would yield a separable mixed state if the measurement outcome is
unknown or discarded.
\bibitem{Footnote6} As above, an idealized situation
in a simplified model is considered. A detailed discussion
including the effects of a finite optical depth and atomic motion
can be found in \cite{EbDtheory,EbDExperiment}.
\bibitem{FootnoteOperatorsDefinition}
The definition of the operators $A$ and $B$ differs from the
definition used in \cite{EbDtheory}, since the atomic states for
the second ensemble are defined differently (here,
$|\!\!\uparrow\rangle_{II}\equiv |4,-3\rangle$ and
$|\!\!\downarrow\rangle_{II}\equiv |4,-4\rangle$, while
$|\!\!\uparrow\rangle_{II}\equiv |4,-4\rangle$ and
$|\!\!\downarrow\rangle_{II}\equiv |4,-3\rangle$ is used in
\cite{EbDtheory}).
\bibitem{Footnote7}
To this end, a sequence of infinitesimal pulses which interact
with the ensembles and are subject to continuous measurements is
considered. By averaging over the measurement results, the master
equation is obtained.
\bibitem{FootnoteMirror}
We assume the canonical conjugate operators for the mechanical
system to be dimensionless $[x_m,p_m]=i$.
\end{thebibliography}
\end{document}